\documentclass[11pt]{article}
\pdfoutput=1

\usepackage{graphics, color,soul}
\usepackage{graphicx}
\usepackage{amssymb}

\usepackage{lmodern,mathtools}

\usepackage{booktabs}
\usepackage[english]{babel}
\usepackage{amsmath,amssymb,amsbsy,amstext, amsthm, simplewick}
\usepackage{hyperref}
\usepackage{tikz}

\usetikzlibrary{decorations.pathmorphing}
\tikzset{snake it/.style={decorate, decoration=snake}}

\usepackage{amsfonts}
\usepackage{amssymb}
\usepackage{upgreek}
\usepackage{simplewick}
 \usepackage{exscale,relsize}
\usepackage{mathtools}
\usepackage{comment}

\usepackage[margin=1cm,labelfont={sf,bf,scriptsize},textfont={sf,scriptsize}]{caption}

\usepackage{colortbl}
\definecolor{lightgreen}{cmyk}{0.2, 0, 0.2, 0.2}
\definecolor{lightgray}{cmyk}{0.1,0.2,0,0.1}
\definecolor{lightgray2}{cmyk}{0.1,0.1,0,0.1}

\makeatletter
\newlength{\apb@width}
\newcommand{\autoparbox}[2][c]{\settowidth{\apb@width}{#2}\parbox[#1]{\apb@width}{#2}}

\makeatother

\numberwithin{equation}{section}

\def\beq{\begin{equation}}
\def\eeq{\end{equation}}

\def\bea{\begin{eqnarray}}
\def\eea{\end{eqnarray}}

\def\beq{\begin{equation}}
\def\eeq{\end{equation}}
\def\be{\begin{equation}}
\def\ee{\end{equation}}
\def\bea{\begin{eqnarray}}
\def\eea{\end{eqnarray}}

\def\0{{\vec{0}}}

\DeclareRobustCommand{\SkipTocEntry}[4]{}

\def\beq{\begin{equation}}
\def\eeq{\end{equation}}

\def\ba#1\ea{\begin{align}#1\end{align}}
\def\bg#1\eg{\begin{gather}#1\end{gather}}
\newcommand{\bseq}{\begin{subequations}}
\newcommand{\eseq}{\end{subequations}}

\DeclareSymbolFont{extraup}{U}{zavm}{m}{n}
\DeclareMathSymbol{\varheart}{\mathalpha}{extraup}{86}
\DeclareMathSymbol{\vardiamond}{\mathalpha}{extraup}{87}

\def\({\left(}
\def\){\right)}
\def\[{\left[}
\def\]{\right]}

\begin{document}

\begin{titlepage}

\setcounter{page}{1} \baselineskip=15.5pt \thispagestyle{empty}

\vbox{\baselineskip14pt
}
{~~~~~~~~~~~~~~~~~~~~~~~~~~~~~~~~~~~~
~~~~~~~~~~~~~~~~~~~~~~~~~~~~~~~~~~
~~~~~~~~~~~ }

\bigskip\

\vspace{2cm}
\begin{center}
{\fontsize{19}{36}\selectfont  \sc
The Dangerous Irrelevance of String Theory
}
\end{center}

\vspace{0.6cm}

\begin{center}
{\fontsize{13}{30}\selectfont  Eva Silverstein}
\end{center}


\begin{center}
\vskip 8pt

\textsl{
\emph{Stanford Institute for Theoretical Physics and Department of Physics, Stanford University, Stanford, CA 94305}}



\vskip 7pt
\textsl{ \emph{Kavli Institute for Particle Astrophysics and Cosmology, Stanford, CA 94025}}

\end{center}

\vspace{0.5cm}
\hrule \vspace{0.1cm}
{ \noindent \textbf{Abstract}

We comment on the relation between string theory and empirical science, grounding our discussion in cosmology, a subject with increasingly precise data in which this connection operates at several levels.  It is important to take into account the phenomenon of dangerous irrelevance:  over long times or large field ranges, physics can become sensitive to higher scales than the input energies.  This pertains in inflationary cosmology (and possibly other aspects of horizon physics).   String theory also contributes to our understanding of observational constraints and search strategies at the level of low energy field theory.  We illustrate this with a current example concerning a new form of non-Gaussianity generated by very massive degrees of freedom coupling to the inflaton.  New constraints on such fields and couplings can be obtained from existing data, increasing our empirical knowledge of the universe.     
This builds in part from the development of the string landscape, which is neither random nor an abdication of science as has sometimes been suggested. {\it Invited contribution to the proceedings of the conference `Why trust a theory'.}
 
\vspace{0.3cm}
\hrule
\vspace{0.6cm}}
\end{titlepage}

\tableofcontents

\section{Introduction}

I was asked by R. Dawid to provide a perspective for the proceedings of
the meeting `Why trust a theory'.  I did not participate in this meeting, but will be happy to comment, focusing on aspects not emphasized in other contributions.   Both sides of the debate \cite{EllisSilk}\cite{Dawid}\ start from the assumption that string theory is divorced from empirical observation.  In this note, I will describe a concrete role that string theory has been playing in the standard scientific method, in the context of early universe cosmology \cite{inflation, Planckpapers, WMAP, SPT, ACT}.  Along the way I will make some comments on the nature and role of the string landscape.  

\subsection{Dangerous Irrelevance and the precision of modern data}

Effective field theory provides a powerful method of characterizing observables, incorporating the limitations of accessible energy scales.  
In assessing the applicability of the physics of the `UV' completion, whatever it is, one must understand carefully the scope of the low energy theory.  
Obviously the energy scale of new physics (e.g. masses of additional fields) may be substantially higher than those excited by terrestrial experiments or observable processes
in the universe, leading to effects suppressed by this ratio\footnote{This was reviewed in another contribution to this volume \cite{JoeTrust}, while \cite{fernandotrust}\ describes various interesting phenomenological scenarios in string theory.}.  
One expresses this in terms of an effective action.  For a scalar field $\phi$, this is schematically
\begin{eqnarray}\label{EFTac}
S=S_{kin}-\int d^4 x \sqrt{-g}\{\frac{1}{2}m^2(\phi &-&\phi_0)^2+\lambda_1(\phi-\phi_0)+\lambda_3(\phi-\phi_0)^3 +\lambda_4(\phi-\phi_0)^4 \nonumber \\
 &+& \lambda_6\frac{(\phi-\phi_0)^6}{M_*^2}+\lambda_{4,4}\frac{(\partial\phi)^4}{M_*^4}+\dots \}
\end{eqnarray}
with effective dimensionless couplings for the irrelevant operators such as $\bar\lambda_6=\frac{E^2}{M_*^2}$ decreasing at low energies.  This makes it more difficult to access their effects at lower energies.

However, there are two standard caveats to this which are relevant to the connection between string theory and
observations.  

One is known as {\it dangerous irrelevance}.  An irrelevant operator can become important upon RG flow, or over long timescales, large field ranges or large parameter variations.   Even if the input energy density is
low, a system may develop sensitivity to the UV completion.  In the above action, this can be seen for example in the potential energy terms:  if
the field travels a distance $\ge M_*$, then we cannot neglect the higher order terms.    This applies in inflationary cosmology  \cite{inflationreviews, TASIlectures}, a point made forcefully in \cite{KKLMMT}.
A more basic physical example of this is to consider charges in a weak electric field\footnote{We thank S. Hartnoll and B. Swingle for discussion of such simple examples.} : over
sufficiently long times, the charges will be accelerated, developing a much larger center of mass energy.  A similar effect occurs in black hole geometries \cite{BHpaper}.

The second caveat has to do with the amount of available data.  Roughly speaking, with $N_m$ measurements, one can constrain a parameter $\epsilon$
down to
\be\label{epsN}
\Delta\epsilon \sim \frac{1}{\sqrt{N_m}}
\ee
With sufficiently many data points, this can lead to sensitivity to mass scales above those directly excited.  
This has sufficed to rule out proton decay in classic GUT models \cite{GUTconstraint}.  
In current cosmology,
this number is approximately $N_m\sim 10^6$  \cite{Planckpapers, WMAP, SPT, ACT}\ (and growing), in a subject where the input energy densities can be rather large to begin with, and
the timescales also imply UV sensitivity \cite{inflationreviews, TASIlectures}\ as just described.   

Although the data is sufficient to constrain a variety of interesting possibilities for new physics, one could use up the data by searching
for all possibilities allowed in an EFT (for example, in inflationary cosmology this would
occur without strong symmetry assumptions, even at the level of single-field inflation \cite{EFTinflation}).  For this reason as well, UV-complete physics can play a useful part 
in suggesting, characterizing and prioritizing analyses.    I should stress that here I am alluding to many possibilities for effects that are subleading compared to the basic, empirically tested, features of inflationary cosmology \cite{Planckpapers, WMAP, SPT, ACT}.\footnote{These include the shape of the temperature and E mode power spectrum, indicating super-horizon perturbations \cite{superhorizon} and a small tilt, as well as the roughy Gaussian shape of the histogram of temperature fluctuations, along with more precise constraints on several shapes of non-Gaussianity.} 

Here I do not mean to claim that the constraints from UV completion will turn out to be strong enough to whittle down the testable possibilities enough to avoid the problem of the look elsewhere effect.   But it can help make sense of the observational constraints in ways that would not follow from EFT alone.  One of my favorite examples of this is qualitative feature that even very massive (i.e. UV) degrees of freedom tend to adjust in an energetically favorable way to flatten the inflaton potential, driving down predictions for the tensor to scalar ratio compared to the corresponding models without such fields \cite{flattening}.  More generally, given the impossibility of testing all the parameters in the EFT, we can reduce the problem in several ways:  (i) impose extra symmetry and/or minimal field content in the EFT\cite{EFTinflation},  (ii) test specific mechanisms which involve the interesting structures of the UV completion, and (iii) do less specific (and hence less optimal) tests of qualitative features suggested by theoretical mechanisms.  This combination leads to interesting constraints on early universe physics; I will give a current example illustrating this below. 

\subsection{Spinoffs and systematics of effective field theory and data constraints}    

In a nutshell:  a basic role string theory plays is in its spinoffs for effective field theory and data analysis.  The theory has
stimulated numerous interesting phenomenological ideas worth testing.  One classic example is low-energy supersymmetry.  This symmetry is actually rare in
the string landscape, but a beautiful idea to test regardless.  Further examples have emerged repeatedly in early universe cosmology.  In addition to being tested in their own right, these models again helped to stimulate a more complete EFT treatment of inflation and its signatures.

\subsection{Timescales, null results, and information theory}

It is sometimes said that theory has strayed too far from experiment/observation.  Historically, there are classic cases
with long time delays between theory and experiment -- Maxwell's and Einstein's waves being prime examples, at 25 and 100 years
respectively\footnote{as well as plenty of delays going the other way (e.g. ... years and counting from the discovery
of high-$T_C$ superconductors)}. These are also good examples of how theory is constrained by serious mathematical and thought-experimental consistency conditions.   

Of course electromagnetism and general relativity are not representative of most theoretical ideas, but the point remains
valid.  When it comes to the vast theory space being explored now, most testable ideas will be constrained or falsified.   Even there
I believe there is substantial scientific value to this:   we learn something significant by ruling out a valid theoretical possibility, as long
as it is internally consistent and interesting.   We also learn important lessons in excluding potential alternative theories based on
theoretical consistency criteria.  The pursuit of no go theorems -- and conversely the exceptions to their assumptions -- is a standard and often useful part of theoretical physics research.    Whether empirical or mathematical, constraints on interesting regions of theory space is valuable science.  In this note we focus on string theory's role in the former.  

Since information theory is currently all the rage, it occurred to me that we can phrase this in that language.  Information is
maximized when the probabilities are equal for a set of outcomes, since one learns the most from a measurement in that case.
The existence of multiple consistent theoretical possibilities implies greater information content in the measurements.
Therefore, theoretical research establishing this (or constraining the possibilities) is directly relevant to the question of what and
how much is learned from data.   In certain areas, string theory plays a direct role in this process.

One thing that is certainly irrelevant to these questions is the human lifespan.  Arguments of the sort `after X number of
years, string theory failed to produce Y result' are vacuous.  In any case, 
we are fortunate that the timescales for testing certain ideas are not so long.  

\subsection{The many facets of string theory}

Before going on, let us note that string theory has many motivations, developments, and potential applications.  It is a strong candidate for a
consistent UV completion of gravity.  It is a rich framework
for physical model-building (in high energy physics and condensed matter as well as cosmology),  a detailed source of mathematical ideas, an approach to black hole physics and other thought-experimental puzzles,
and a source of insight into quantum field theory.  In some ways its effectiveness is `unreasonable', although in other ways it has not developed the way some expected/hoped.  
With all this, it would be difficult to justify any simple ideological stand on the justification of string theory as a subfield, despite the tendencies
of internet discourse to try to do so.  Anyway 
in this note we are focused on an important, but narrower, question:  how and why does string theory participate in current empirical science?  

\section{Case study:  early universe cosmology}

\subsection{Empirical Observations}

Recent decades have seen enormous progress in our understanding of cosmological evolution and the physical
processes involved, raising deeper questions and stimulating further observations.  
In cosmology, as in many areas of science, we would like to optimally exploit the available data, especially given
the enormous ingenuity and effort that goes into collecting and analyzing it \cite{Planckpapers, WMAP, SPT, ACT}.  The first step
in this enterprise, which has nothing directly to do with string theory, is to nail down the known cosmological parameters
with greater accuracy and precision.  This is a dry way of stating an enormously interesting process which has
led, among many other things, to the discovery of accelerated expansion of the universe.  

On top of that, we can use the data to test for new parameters and to interpret the resulting constraints, and it is in this latter process that string theory already plays a significant role.   
There are by now many examples of this (see reviews such as \cite{inflationreviews, TASIlectures}\ for references), including examples such as DBI inflation which broadened our understanding of inflationary dynamics while generating a now-standard shape of non-Gaussianity, and large-field inflation mechanisms relevant for B mode searches (as well as more model dependent structures in the scalar perturbations).  We will not rehash these, but make some brief comments and then spell out a relatively new example for illustration.    

Additional fields, including very heavy ones, that are coupled to the inflaton have numerous effects, some detectable/constrainable observationally.  As mentioned above, they adjust to the inflationary potential in an energetically favorable way, either destabilizing or flattening the inflaton potential depending on the details \cite{flattening}.\footnote{The latter effect accounts for the continued viability of string-theoretic inflationary mechanisms as of this writing.} They renormalize the effective action for the inflaton, and they are subject to non-adiabatic production, slowing the homogenous rolling inflaton and leading to Bremsstrahlung emission of scalar and tensor perturbations.\footnote{They also feature in an interesting proposal for interpreting the low-multipole anomalies \cite{Bondetalanomalies}.}    Several of these basic effects were discovered in the context of string theoretic inflationary mechanisms, from which more model-independent lessons were abstracted.  

\subsubsection{A current example for illustration}

Given an inflationary solution, the perturbations can be measurably affected by even heavy fields $\chi$ that couple to the inflaton.  Such couplings produce a time-dependent mass.  Consider the dominant Fourier component of this mass function, an interaction $\chi^2(\mu^2+g^2 f^2\cos\frac{\phi}{f})$.  Expanding the inflaton into its background evolution and perturbations, $\phi=\dot\phi_0 t+\delta\phi$ generates a series of interaction terms, each of which has a sinusoidal time dependence at a frequency $\omega=\frac{\dot\phi_0}{f}$.  For example, there is a 3-point vertex $\int g^2 f\chi^2\delta\phi\sin\omega t$. This causes non-adiabatic production that is exponentially suppressed, but only beyond the mass scale $\sim g\dot\phi$ (and the amplitude of these effects are enhanced by power law prefactors \cite{productive}).  This, combined with the precision of the CMB, leads to sensitivity to mass scales up to two orders of magnitude above the Hubble scale of inflation, for  a range of couplings consistent with perturbation theory.

The three point correlation function of scalar perturbations is of a distinct shape from those previously analyzed in the data (as well as in theory).   
Moreover, there is a regime in which this effect has a strongly non-Gaussian shape, in contrast to all previously derived signatures.  The $\sin\omega t$ dependent coupling enhances each leg of $\delta\phi$ emission from a given pair of produced $\chi$ particles.  As a result, a regime emerges for $\frac{\omega}{H}\gg 1$ in which the signal/noise ratio grows with $N$ for a range of $N$.  
Writing this in terms of the more directly observed perturbation $\zeta \sim \delta\phi H/\dot\phi$, this ratio is schematically
\be\label{SNratio}
(S/N)^2\sim \int_{\{ k \}} \frac{|\langle\zeta_1\dots\zeta_N\rangle|^2}{N! \prod P(k_i)}\sim \frac{\left(g^2\frac{f}{\mu}\sqrt{\frac{\omega}{H}}\right)^N}{N!}    
\ee
in the regime where the Gaussian perturbations dominate the noise in the denominator.  If one derives a histogram of temperature fluctuations generated by this effect, it is a strongly non-Gaussian shape in $\zeta$ space, convolved with a Gaussian distribution arising from the standard vacuum fluctuations of $\zeta$.  One can derive the effect this would have on the CMB map in position space:  it gives a nearly scale-invariant pattern of defects of different sizes, with $\delta\phi$ perturbations radiating out from each $\chi$ production point, distorting the map within the light cone of each such event.  
This leads to several new types of non-Gaussianity searches underway using CMB data.  

What does this have to do with string theory?  There are several levels.  

$\bullet$  First, this theoretical analysis grew out of a class of mechanisms for inflation in string theory (known as monodromy and trapped inflation \cite{inflationreviews}), where the ingredients leading to the effect arise naturally.  This includes heavy fields coupling to the inflaton, in some cases sinusoidal couplings for string theory's analogue of axion fields.  As such, the data analysis will put constraints on the parameter space of models for inflation in this class.  


$\bullet$ Second, we can abstract from this the broader lesson that for a range of interesting masses and couplings, we must supplement the single-field EFT even if the extra particles are very heavy, and their effects can be constrained using existing CMB data \cite{Moritzetal}.   (This lesson also applies to numerous other string-theoretic inflationary models, which also contain heavy fields.)    In ongoing theoretical work, we are currently finding other roles for large-$N$ point functions in non-Gaussianity, arising from combinatorial enhancements of such observables.   

$\bullet$ At a third level, the theoretical demonstration that strongly non-Gaussian effects can arise in the promordial perturbations motivates a more extensive analysis of the theory and analysis of strongly non-Gaussian perturbations, which can have other origins.    

$\bullet$  A similar effect arises from time dependent string tensions, with a related search strategy. 
  
The middle two bullet points are logically independent of string theory, but as has happened several times in the past, we came to them via string theory (as in the first bullet point).   The connection between string theory and early universe cosmology, which has been very active across the spectrum from theory to data analysis, was not covered in \cite{EllisSilk}.  

        
I should emphasize that these developments derive from the `landscape' of string theory solutions \cite{BoussoPolchinski, landscapereviews}, which I will discuss further below.  Rather than representing an abdication of science, or randomizing all physical observables, the landscape has led to new empirical information about the early universe as well as providing a consistent interpretation of the dark energy.  

\subsection{Thought Experiments}

The role of thought experiments is important as well, needless to say.  In the context of cosmology, one role these play is to constrain possible alternative scenarios for the initial perturbations, which is an interesting line of research.   Some prominent examples can be excluded this way if one requires that black hole thermodynamics relations hold.  Exotic forms of stress-energy required to introduce a bounce in the cosmological scale factor can violate these relations.  In such cases, this leads to decreasing black hole mass as entropy increases, violating the second law.

\section{Demystifying the landscape}

It is sometimes said that the landscape makes every parameter into an unexplainable selection effect.  But even though there are many backgrounds of string theory, the landscape is still highly constrained.  To begin with, there is not a hard cosmological constant term in the effective Lagrangian descending from string theory.  The resulting metastability of de Sitter vacua fits with several conceptual (thought-experimental) and technical arguments.  It fits well with the observation of the dark energy.  
The middle ground between structure and variability complicates the task of modeling cosmology and particle physics in explicit detail, although as already noted it inspires mechanisms and dynamical effects testable in their own right.   This program of research remains highly motivated, with new discoveries continuing to emerge in both directions: structures and mechanisms, and constraints.  

\subsection{The role of supersymmetry}

In my view,  the role of supersymmetry is chronically over-emphasized in the field, and hence understandably also in the article by Ellis and Silk \cite{EllisSilk}.    The possibility of supersymmetry in nature is very interesting since it could stabilize the electroweak hierarchy, and extended supersymmetry enables controlled extrapolation to strong coupling in appropriate circumstances.  Neither of these facts implies that low-energy supersymmetry is phenomenologically favored in string theory.  

Almost every perturbative string limit has a positive dilaton potential, and almost every compactification geometry is negatively curved leading to a positive potential from the internal curvature.   The elegant mathematics of string theory applies to these cases too; for example there is a generalization of T-duality that neatly relates the two; compact negatively curved spaces are supercritical as a result of their exponentially growing fundamental group \cite{Dduality}.   To date, most works on string compactification presumes the extra dimensions to be six in number and built from a Ricci-flat (Calabi-Yau) manifold, chosen by hand to preseve supersymmetry below the compactification scale.  Other, less supersymmetric mechanisms with positive tree level potential have also been studied, and ongoing work is uncovering interesting structure, and some simplifications, in the more generic setting.  It is important to note that perturbative control at large curvature radius and weak string coupling is available in both cases.  Much further research, both conceptual and technical, is required to obtain an accurate assessment of the dominant contributions to the string landscape.        

\subsection{Is it testable as a whole?}

The discussion above and more comprehensively in \cite{TASIlectures}\ emphasized the testability of certain inflationary mechanisms descending from string theory, in modeling a process that ideally require a quantum gravitational treatment (as well as the more nuanced role these play in the interpretation of empirical observations).  A somewhat separate question is the testability of the landscape itself, leading to some of the more philosophical discussion in these proceedings.  But the two questions are not necessarily distinct.  In principle one could test string theory locally.  In practice, this would require discovering a smoking gun signature (such as a low string scale at colliders, or perhaps a very distinctive pattern of primordial perturbations in cosmology), and nothing particularly favors such scenarios currently.  But for  the philosophical question of the empirical standing of the string landscape, this is an important point to include.   Strong evidence for string theory locally would             
support its global predictions of a landscape.\footnote{As my colleague A. Linde frequently points out, the universe hypothesis is no more conservative than the multiverse hypothesis in the sense that both refer to physics outside of our empirical view.}  

This is an extreme example of a familiar chain of reasoning in science.  Even in empirically established theories, we empirically test only a set of measure zero of their predictions.  A plethora of such tests can provide compelling evidence for a theory, which makes further predictions beyond those explicitly tested.  It is not ever the case that {\it all} of a theory's predictions are empirically verified.    
In any case, it is reasonable to test string theoretic physical models locally as far as possible, while continuing to assess the theory's implications more globally.



\section{Summary}

String theory participates in empirical science in several ways.  
In the context of early universe cosmology, on which we have focused in this article, it helped motivate the discovery and development of mechanisms for dark energy and inflation consistent with the mathematical structure of string theory and various thought-experimental constraints.  Some of these basic mechanisms had not been considered at all outside of string theory, and some not quite in the form they take there, with implications for effective field theory and data analysis that go well beyond their specifics.  Low-energy supersymmetry -- a very special choice within the string landscape --is an earlier example of a major idea originating in string theory which is well worth testing although it is not a general prediction of string theory as far as we know.  A subset of models originating in string theory generate rich signatures, enabling direct constraints on their parameters, with some falsifiable.  A current example concerns an entirely new form of non-Gaussianity generated by very massive degrees of freedom in the early universe.  As in previous examples, new constraints on these can be obtained from existing data, increasing our empirical knowledge of the universe via the standard scientific method, regardless of whether the analysis results in a discovery or a null result.     

This is an active area in which big open questions remain both theoretically and observationally, and it is too soon to draw conclusions about the ultimate level of empirical connection that string theory will attain.  In any case, it already plays a useful role in interpreting empirical observations of significant interest.   
 
\section*{Acknowledgements}

I thank many collaborators and colleagues for sharing their insights over many years, and J. Polchinski for commenting on a draft.  I would like to thank L. Senatore and KIPAC at Stanford for an early discussion addressing the claims and proposals in \cite{EllisSilk}.  
The scientific research described here is supported in part by the National Science Foundation
under grant PHY-0756174.

\newpage



\begingroup\raggedright\begin{thebibliography}{10}
\baselineskip=14.5pt

\bibitem{EllisSilk}

G. Ellis and J. Silk,
``Scientific method: Defend the integrity of physics'',
Nature, Volume 516, Issue 7531, pp. 321-323 (2014).

\bibitem{Dawid}

R. Dawid,
``Non-empirical Confirmation'',
American Astronomical Society, AAS Meeting number 228, id.211.02

\bibitem{inflation}
  A.~H.~Guth,
  ``The Inflationary Universe: A Possible Solution to the Horizon and Flatness Problems,''
  Phys.\ Rev.\ D {\bf 23}, 347 (1981).

  A.~D.~Linde,
  ``A New Inflationary Universe Scenario: A Possible Solution of the Horizon, Flatness, Homogeneity, Isotropy and Primordial Monopole Problems,''
  Phys.\ Lett.\ B {\bf 108}, 389 (1982).

  A.~Albrecht and P.~J.~Steinhardt,
  ``Cosmology for Grand Unified Theories with Radiatively Induced Symmetry Breaking,''
  Phys.\ Rev.\ Lett.\  {\bf 48}, 1220 (1982).

 A.~A.~Starobinsky,
  ``A New Type of Isotropic Cosmological Models Without Singularity,''
  Phys.\ Lett.\ B {\bf 91}, 99 (1980).

 \bibitem{Planckpapers}


P.~A.~R.~Ade {\it et al.}  [Planck Collaboration],
  ``Planck 2015 results. XIII. Cosmological parameters,''
  arXiv:1502.01589 [astro-ph.CO].
  
   N.~Aghanim {\it et al.} [Planck Collaboration],
  ``Planck 2015 results. XI. CMB power spectra, likelihoods, and robustness of parameters,''
  Submitted to: Astron.Astrophys.
  [arXiv:1507.02704 [astro-ph.CO]].
  
 P.~A.~R.~Ade {\it et al.} [Planck Collaboration],
  ``Planck 2015 results. XVI. Isotropy and statistics of the CMB,''
  arXiv:1506.07135 [astro-ph.CO].


P.~A.~R.~Ade {\it et al.}  [Planck Collaboration],
  ``Planck 2015 results. XVII. Constraints on primordial non-Gaussianity,''
  arXiv:1502.01592 [astro-ph.CO].


P.~A.~R.~Ade {\it et al.}  [BICEP2 and Planck Collaborations],
  ``Joint Analysis of BICEP2/$Keck  Array$ and $Planck$ Data,''
  Phys.\ Rev.\ Lett.\  {\bf 114}, no. 10, 101301 (2015)
  [arXiv:1502.00612 [astro-ph.CO]].


  P.~A.~R.~Ade {\it et al.}  [Planck Collaboration],
  ``Planck 2015 results. XX. Constraints on inflation,''
  arXiv:1502.02114 [astro-ph.CO].

\bibitem{WMAP}

G.~Hinshaw {\it et al.}  [WMAP Collaboration],
  ``Nine-Year Wilkinson Microwave Anisotropy Probe (WMAP) Observations: Cosmological Parameter Results,''
  Astrophys.\ J.\ Suppl.\  {\bf 208}, 19 (2013)
  [arXiv:1212.5226 [astro-ph.CO]].

\bibitem{SPT}

K.~T.~Story, C.~L.~Reichardt, Z.~Hou, R.~Keisler, K.~A.~Aird, B.~A.~Benson, L.~E.~Bleem and J.~E.~Carlstrom {\it et al.},
  ``A Measurement of the Cosmic Microwave Background Damping Tail from the 2500-square-degree SPT-SZ survey,''
  arXiv:1210.7231 [astro-ph.CO].

\bibitem{ACT}

J.~L.~Sievers, R.~A.~Hlozek, M.~R.~Nolta, V.~Acquaviva, G.~E.~Addison, P.~A.~R.~Ade, P.~Aguirre and M.~Amiri {\it et al.},
  ``The Atacama Cosmology Telescope: Cosmological parameters from three seasons of data,''
  arXiv:1301.0824 [astro-ph.CO].
  
\bibitem{JoeTrust}

 J.~Polchinski,
  ``String theory to the rescue,''
  arXiv:1512.02477 [hep-th].

\bibitem{fernandotrust}

F.~Quevedo,
  ``Is String Phenomenology an Oxymoron?,''
  arXiv:1612.01569 [hep-th].

\bibitem{GUTconstraint}

K.~Abe {\it et al.} [Super-Kamiokande Collaboration],
  ``Search for proton decay via $p \to e^+\pi^0$ and $p \to \mu^+\pi^0$ in 0.31  megaton·years exposure of the Super-Kamiokande water Cherenkov detector,''
  Phys.\ Rev.\ D {\bf 95}, no. 1, 012004 (2017)
  doi:10.1103/PhysRevD.95.012004
  [arXiv:1610.03597 [hep-ex]].
    
\bibitem{inflationreviews}

 D.~Baumann and L.~McAllister,
  ``Inflation and String Theory,''
  arXiv:1404.2601 [hep-th].

D.~Baumann and L.~McAllister,
  ``Advances in Inflation in String Theory,''
  Ann.\ Rev.\ Nucl.\ Part.\ Sci.\  {\bf 59}, 67 (2009)
  [arXiv:0901.0265 [hep-th]].

D.~Baumann,
  ``TASI Lectures on Inflation,''
  arXiv:0907.5424 [hep-th].
  
  C.~P.~Burgess and L.~McAllister,
  ``Challenges for String Cosmology,''
  Class.\ Quant.\ Grav.\  {\bf 28}, 204002 (2011)
  doi:10.1088/0264-9381/28/20/204002
  [arXiv:1108.2660 [hep-th]].
  
   A.~Westphal,
  ``String Cosmology -- Large-Field Inflation in String Theory,''
  Adv.\ Ser.\ Direct.\ High Energy Phys.\  {\bf 22}, 351 (2015).
  $doi:10.1142/9789814602686_0012$

\bibitem{TASIlectures}

 E.~Silverstein,
  ``TASI lectures on cosmological observables and string theory,''
 in {\it Proceedings, Theoretical Advanced Study Institute in Elementary Particle Physics: New Frontiers in Fields and Strings (TASI 2015)} : Boulder, CO, USA, June 1-26, 2015 

  arXiv:1606.03640 [hep-th].
  
 \bibitem{KKLMMT}  
      
S.~Kachru, R.~Kallosh, A.~D.~Linde, J.~M.~Maldacena, L.~P.~McAllister and S.~P.~Trivedi,
  ``Towards inflation in string theory,''
  JCAP {\bf 0310}, 013 (2003)
  doi:10.1088/1475-7516/2003/10/013
  [hep-th/0308055].
  
\bibitem{BHpaper}

L.~Susskind,
  ``Strings, black holes and Lorentz contraction,''
  Phys.\ Rev.\ D {\bf 49}, 6606 (1994)
  doi:10.1103/PhysRevD.49.6606
  [hep-th/9308139].

  M.~Dodelson and E.~Silverstein,
  ``String-theoretic breakdown of effective field theory near black hole horizons,''
  arXiv:1504.05536 [hep-th].
  
 D.~A.~Lowe, J.~Polchinski, L.~Susskind, L.~Thorlacius and J.~Uglum,
  ``Black hole complementarity versus locality,''
  Phys.\ Rev.\ D {\bf 52}, 6997 (1995)
  doi:10.1103/PhysRevD.52.6997
  [hep-th/9506138].

  R.~Ben-Israel, A.~Giveon, N.~Itzhaki and L.~Liram,
  ``Stringy Horizons and UV/IR Mixing,''
  JHEP {\bf 1511}, 164 (2015)
  doi:10.1007/JHEP11(2015)164
  [arXiv:1506.07323 [hep-th]].

 M.~Dodelson, E.~Silverstein and G.~Torroba,
  ``Varying dilaton as a tracer of classical string interactions,''
  arXiv:1704.02625 [hep-th].

  M.~Dodelson and E.~Silverstein,
  ``Long-Range Nonlocality in Six-Point String Scattering: simulation of black hole infallers,''
  arXiv:1703.10147 [hep-th].

E.~Silverstein,
  ``Backdraft: String Creation in an Old Schwarzschild Black Hole,''
  arXiv:1402.1486 [hep-th].

  A.~Puhm, F.~Rojas and T.~Ugajin,
  ``(Non-adiabatic) string creation on nice slices in Schwarzschild black holes,''
  JHEP {\bf 1704}, 156 (2017)
  doi:10.1007/JHEP04(2017)156
  [arXiv:1609.09510 [hep-th]].

D. Wenren, ``Hyperbolic Black Holes and Open String Production'', to appear.

M. Dodelson et al, ``Long-range longitudinal scattering in holographic
gauge theory and black holes'', to appear.  
    
 \bibitem{EFTinflation}
 
  C.~Cheung, P.~Creminelli, A.~L.~Fitzpatrick, J.~Kaplan and L.~Senatore,
  ``The Effective Field Theory of Inflation,''
  JHEP {\bf 0803}, 014 (2008)
  doi:10.1088/1126-6708/2008/03/014
  [arXiv:0709.0293 [hep-th]].

 \bibitem{superhorizon}
 
 D.~N.~Spergel and M.~Zaldarriaga,
  ``CMB polarization as a direct test of inflation,''
  Phys.\ Rev.\ Lett.\  {\bf 79}, 2180 (1997)
  doi:10.1103/PhysRevLett.79.2180
  [astro-ph/9705182].

N.~Turok,
  ``A Causal source which mimics inflation,''
  Phys.\ Rev.\ Lett.\  {\bf 77}, 4138 (1996)
  doi:10.1103/PhysRevLett.77.4138
  [astro-ph/9607109].

\bibitem{flattening}

X.~Dong, B.~Horn, E.~Silverstein and A.~Westphal,
  ``Simple exercises to flatten your potential,''
  Phys.\ Rev.\ D {\bf 84}, 026011 (2011)
  doi:10.1103/PhysRevD.84.026011
  [arXiv:1011.4521 [hep-th]].

\bibitem{BoussoPolchinski}

 R.~Bousso and J.~Polchinski,
  ``Quantization of four form fluxes and dynamical neutralization of the cosmological constant,''
  JHEP {\bf 0006}, 006 (2000)
  doi:10.1088/1126-6708/2000/06/006
  [hep-th/0004134].
 
  A.~Maloney, E.~Silverstein and A.~Strominger,
  ``De Sitter space in noncritical string theory,''
  hep-th/0205316.
  
 S.~B.~Giddings, S.~Kachru and J.~Polchinski,
  ``Hierarchies from fluxes in string compactifications,''
  Phys.\ Rev.\ D {\bf 66}, 106006 (2002)
  doi:10.1103/PhysRevD.66.106006
  [hep-th/0105097].
  
S.~Kachru, R.~Kallosh, A.~D.~Linde and S.~P.~Trivedi,
  ``De Sitter vacua in string theory,''
  Phys.\ Rev.\ D {\bf 68}, 046005 (2003)
  doi:10.1103/PhysRevD.68.046005
  [hep-th/0301240].

\bibitem{landscapereviews}

A.~R.~Frey,
  ``Warped strings: Selfdual flux and contemporary compactifications,''
  hep-th/0308156.

J.~Polchinski,
  ``The Cosmological Constant and the String Landscape,''
  hep-th/0603249.

M.~R.~Douglas and S.~Kachru,
  ``Flux compactification,''
  Rev.\ Mod.\ Phys.\  {\bf 79}, 733 (2007)
  [hep-th/0610102].

 F.~Denef,
  ``Les Houches Lectures on Constructing String Vacua,''
  arXiv:0803.1194 [hep-th].
  
 E.~Silverstein,
  ``TASI / PiTP / ISS lectures on moduli and microphysics,''
  hep-th/0405068.

\bibitem{Bondetalanomalies}

J.~R.~Bond, A.~V.~Frolov, Z.~Huang and L.~Kofman,
  ``Non-Gaussian Spikes from Chaotic Billiards in Inflation Preheating,''
  Phys.\ Rev.\ Lett.\  {\bf 103}, 071301 (2009)
  doi:10.1103/PhysRevLett.103.071301
  [arXiv:0903.3407 [astro-ph.CO]].

\bibitem{productive}
 R.~Flauger, M.~Mirbabayi, L.~Senatore and E.~Silverstein,
  ``Productive Interactions: heavy particles and non-Gaussianity,''
  arXiv:1606.00513 [hep-th].

\bibitem{Moritzetal}

M. Mirbabayi, M. Munchmeyer et al, work in progress.


\bibitem{Dduality}

 D.~R.~Green, A.~Lawrence, J.~McGreevy, D.~R.~Morrison and E.~Silverstein,
  ``Dimensional duality,''
  Phys.\ Rev.\ D {\bf 76}, 066004 (2007)
  doi:10.1103/PhysRevD.76.066004
  [arXiv:0705.0550 [hep-th]].

  J.~McGreevy, E.~Silverstein and D.~Starr,
  ``New dimensions for wound strings: The Modular transformation of geometry to topology,''
  Phys.\ Rev.\ D {\bf 75}, 044025 (2007)
  doi:10.1103/PhysRevD.75.044025
  [hep-th/0612121].

\endgroup
\end{document}